\documentclass[prl,twocolumn,superscriptaddress,showpacs]{revtex4}
\usepackage[dvips]{graphicx}
\usepackage{color}
\usepackage{amsmath}

\begin{document}

\preprint{LAUR 05-9176}
\title{Spin noise spectroscopy to probe quantum
states of ultracold fermionic atomic gases}
\author{Bogdan~Mihaila}
\affiliation{Theoretical Division,
   Los Alamos National Laboratory,
   Los Alamos, NM 87545}
\author{Scott~A.~Crooker}
\affiliation{National High Magnetic Field Laboratory,
   Los Alamos National Laboratory,
   Los Alamos, NM 87545}
\author{Krastan~B.~Blagoev}
\affiliation{Theoretical Division,
   Los Alamos National Laboratory,
   Los Alamos, NM 87545}
\author{Dwight~G.~Rickel}
\affiliation{National High Magnetic Field Laboratory,
   Los Alamos National Laboratory,
   Los Alamos, NM 87545}
\author{Peter~B.~Littlewood}
\affiliation{Cavendish Laboratory,
   Madingley Road,
   Cambridge CB3 0HE,
   United Kingdom}
\author{Darryl~L.~Smith}
\affiliation{Theoretical Division,
   Los Alamos National Laboratory,
   Los Alamos, NM 87545}




\begin{abstract}
Ultracold alkali atoms~\cite{anderson,davis,bradley,demarco} provide
experimentally accessible model systems for probing quantum states
that manifest themselves at the macroscopic scale.  Recent
experimental realizations of superfluidity in dilute gases of
ultracold fermionic (half-integer spin)
atoms~\cite{regal,bartenstein,zwierlein} offer exciting
opportunities to directly test  theoretical models of related
many-body fermion systems that are inaccessible to experimental
manipulation, such as neutron stars~\cite{pethick} and quark-gluon
plasmas~\cite{qgp}. However, the microscopic interactions between
fermions are potentially quite complex, and experiments in ultracold
gases to date cannot clearly distinguish between the qualitatively
different microscopic models that have been
proposed~\cite{holland,eddy,ref:3level,bruun}. Here, we
theoretically demonstrate that optical measurements of electron spin
noise~\cite{ref:spin,noise} -- the intrinsic, random fluctuations of
spin -- can probe the entangled quantum states of ultracold
fermionic atomic gases and unambiguously reveal the detailed nature
of the interatomic interactions. We show that different models
predict different sets of resonances in the noise spectrum, and once
the correct effective interatomic interaction model is identified,
the line-shapes of the spin noise can be used to constrain this
model. Further, experimental measurements of spin noise in classical
(Boltzmann) alkali vapors are used to estimate the expected signal
magnitudes for spin noise measurements in ultracold atom systems and
to show that these measurements are feasible.
\end{abstract}


\maketitle

Owing to their perceived simplicity, the properties of dilute
degenerate fermionic systems are sometimes attributed universal
character~\cite{unitarity}, an assertion that must be called into
question if the interaction between degenerate alkali atoms turns
out to be complex. It is thus unfortunate that the nature of the
effective interatomic interactions remains an outstanding
fundamental issue. Specifically, the role played by the underlying
hyperfine atomic-level structure is not yet understood. To date,
experiments on ultracold fermionic atom gases can be interpreted
equally well using qualitatively different models for the effective
interactions, such as the Fermi-Bose~\cite{holland,eddy} or the
multi-level models~\cite{ref:3level,bruun}. Hence, new experimental
observables are needed to distinguish between these competing
models, and the principal distinctions between current models will
lie in their predicted excitation spectra, which are not yet well
studied.

The excitation spectra of physical systems are often studied by
measuring their response to an external perturbation. Alternatively,
measuring the spectrum of intrinsic fluctuations of a physical
system can provide the same information, and these ``noise
spectroscopies" often disturb the physical system less strongly and
scale more favorably with system size reduction. At very low
temperature, noise from quantum fluctuations of an observable that
does not commute with the Hamiltonian of the system can be used as a
probe of the system properties. Electron spin is not a good quantum
number in alkali gases, and fluctuations of electron spin can be
measured using optical Faraday rotation.  The electron spin noise
spectrum consists of a series of resonances occurring at frequencies
corresponding to the difference between hyperfine/Zeeman atomic
levels. The integrated strength of the lines gives information about
the occupation of the atomic levels, while the line shapes depend on
the properties of the condensed atomic state. It is precisely the
spectroscopic nature of the electron spin noise measurement that
allows it to distinguish between various many-body models for the
quantum states of ultracold fermionic atom gases: different models
predict entirely different sets of resonances in the noise spectrum,
and once the correct effective interatomic interaction model is
identified, the line-shapes of the spin noise can be used to
constrain this model.

To measure the electron spin noise via Faraday spectroscopy, a
linearly polarized laser beam, with photon energy tuned near but not
exactly on the s-p optical transition of the outer s-orbital
electron, traverses an ensemble of alkali atoms~\cite{ref:spin}. The
rotation angle of the laser polarization traversing the sample is
measured as a function of time.  The time average of this rotation
angle vanishes, but the noise power spectrum of the rotation angle
shows distinct peaks.  In the electronic ground state (s-orbital)
there is a strong hyperfine coupling between the nuclear and
electron spins. For the electronic p-orbital, the hyperfine
splitting is weak, however there is a strong spin-orbit coupling
between the p-orbital and its spin. The laser photons directly
couple to the spatial part of the electron wave function, but
because of the spin-orbit splitting in the final state of the
optical transition there is an indirect coupling between the laser
photons and the electron spin. A fluctuating birefringence, that is
a difference in refractive index for left and right hand circular
polarizations, results from quantum fluctuations in the electron
spin and leads to rotation of the polarization angle of the laser.
The experiment is sensitive to fluctuations of electron spin
projection in the direction of laser propagation.


The Hamiltonian describing the system of alkali atoms consists of
a sum of one- and two-atom terms. The one-atom Hamiltonian
includes the kinetic energy, the Zeeman interaction, and the
hyperfine interaction between the nuclear spin~${\vec {I}}$ and
the electron spin~${\vec {s}}$. The single-atom eigenvectors, the
starting point for describing the many-body system, are obtained
by diagonalizing the one-atom Hamiltonian. This Hamiltonian
preserves the projection of total angular momentum, $\vec F = \vec
s + \vec I$, in the direction of the applied magnetic field. The
one-atom matrix elements can be grouped into 2-dimensional blocks
involving the basis states $|I\, m_I\rangle |\frac{1}{2}
\frac{1}{2}\rangle$ and $|I\, m_I + 1\rangle |\frac{1}{2}
-\frac{1}{2}\rangle$, except for the states $ |I,\, m_I=+I\rangle
|\frac{1}{2} \frac{1}{2}\rangle$, and $|I,\, m_I=-I\rangle
|\frac{1}{2} -\frac{1}{2}\rangle$, which do not couple. The
single-atom energies vary smoothly and do not cross with
increasing magnetic field so the single-atom states can be
labelled unambiguously by, $|F\, m_F\rangle $, where $m_F$ is a
good quantum number, but $F$ is only a good quantum number at zero
magnetic field.

In general, the projections of electron spin are not good quantum
numbers of the one atom Hamiltonian.  At magnetic fields where the
hyperfine interaction is comparable or larger than the Zeeman
splitting, the electron and nuclear spins are entangled and no
projection of electron spin is a good quantum number.  At strong
magnetic fields, where the Zeeman splitting is much larger than the
hyperfine interaction, the electron spin projection in the direction
of the magnetic field becomes a good quantum number, but electron
spin projection orthogonal to the magnetic field is not. Faraday
rotation is sensitive to fluctuations of the electron spin in the
laser propagation direction.  Thus at magnetic fields where the
hyperfine interaction is comparable or larger than the Zeeman
splitting, noise spectroscopy can be performed with the magnetic
field either parallel or orthogonal to the direction of laser
propagation whereas in the opposite limit, the magnetic field must
be orthogonal to the direction of laser propagation.


\begin{figure}[b!]
   \includegraphics[width=0.9\columnwidth]{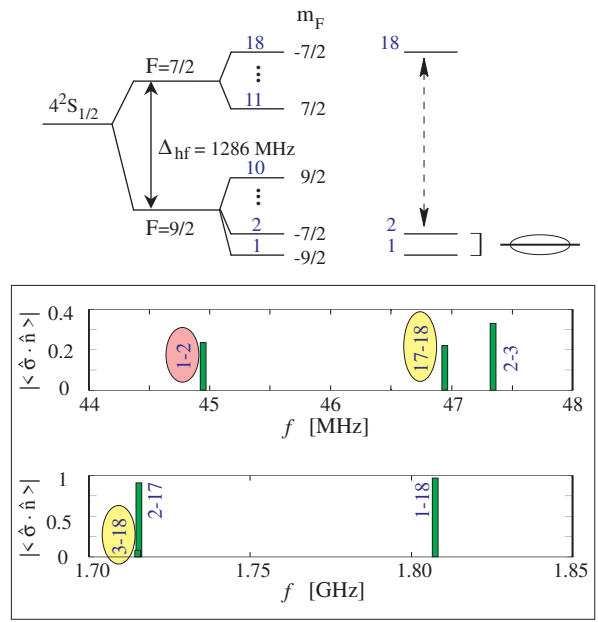}
   \caption{\label{fig:k40_vert}
   [top]
   The Zeeman structure of $^{40}$K (I=4),
   together with schematics of the 3- and 2-level models.
   In the 2-level model, only the two lowest hyperfine states are
   present.
   [bottom]
   Expected spin noise spectrum at the Feshbach resonance ($B=202$ G),
   for laser propagation orthogonal to the magnetic field.
   The highlighted transitions are not present in the 2-level model,
   but are predicted in the 3-level model~\cite{ref:3level}.
   In the 2-level model, lines $17-18$ and $3-18$ are forbidden,
   and line $1-2$ has zero strength.
   Lines $3-18$ and $2-17$ are almost degenerate.
   }
\end{figure}


For a Gaussian optical beam profile, the polarization rotation
angle noise is
\begin{equation}
   \frac{\phi_N(\omega)}{\sqrt{\delta f}} \ = \
   C \
     \Bigl [
       \frac {\sqrt{\pi}}{2} \
       \frac {L \, \rho_0}{A} \
       S(\omega)
     \Bigr ]^{1/2}
   \>,
\label{power}
\end{equation}
where
\begin{equation}
    C \ = \
        \frac{2\pi}{3} \
        \frac{c \, r_0}{m_0} \
        \frac{1}{\hbar \Omega} \
        \frac{ |\langle S|p_x|P_x \rangle| ^2}{|\Omega_{s-p}-\Omega|}
   \>.
\end{equation}
Here $\rho_0$ is the density of atoms, $\Omega$ is the angular
frequency of the laser, $\Omega_{s-p}$ is the angular frequency of
the optical resonance, $r_0$ is the classical electron radius,
$m_0$ is the electron mass, $<S|p_x|P_x>$ is the momentum matrix
element for the optical transition, and the optical beam area is
$A= \pi {q_0}^2$ where $q_0$ is the radius at which the beam
intensity drops to $1/e$ of its peak value. The electron spin
response function $S(\omega)$ is
\begin{equation}
    S(\omega) \ = \
    \frac{1}{\rho_0} \ \int \mathrm{d}t \ e^{i \omega t}
    \int \mathrm{d}^3 r \
    \langle \sigma_z({\bf{r}},t) \sigma_z(0,0) \rangle
   \>,
\end{equation}
where $z$ is the direction of laser propagation,
$\sigma_z({\bf{r}},t)$ is twice the z-projection of the electron
spin density operator, and $S(\omega)$ satisfies the sum rule
\begin{equation}
   \int \mathrm{d}\omega \ S(\omega) \ = \ 2\pi \>.
   \label{sum_rule}
\end{equation}
The spin response function $S(\omega)$ has peaks at
frequencies near the separation between single atom spin levels
\begin{align}
   S(\omega) =
       \sum_{ij} \
       | \langle i | \sigma_z | j \rangle |^2 \
       S^{i \rightarrow j}(\omega)
\label{P_omega_ij}
   \>,
\end{align}
where $\{i,j\}$ label the single atom spin levels $| \langle i |
\sigma_z | j \rangle |^2$ is a one atom matrix element that
determines line strengths and selection rules, and $S^{i \rightarrow
j}(\omega)$ contains information about the many-body state. The
response function $S^{i \rightarrow j}(\omega)$ will only have
strength if at least one of the states $i$ or $j$ has nonzero
occupation.  For laser propagation orthogonal to the magnetic field
the selection rules require that the one-atom quantum number, $m_F$
change by $\pm 1$ between the two single atom levels.

Equations (1) and (2) show that the noise signal decreases linearly
with inverse frequency detuning from the optical resonance. By
contrast, the energy dissipated into the atomic system, either by
optical absorption or Raman scattering, decreases quadratically with
inverse frequency detuning.  Thus noise spectroscopy measurements
are weakly perturbative in the sense that the noise spectroscopy
signal deceases more slowly with inverse frequency detuning than
does the energy dissipated into the system.


\begin{figure}[t!]
   \includegraphics[width=0.95\columnwidth]{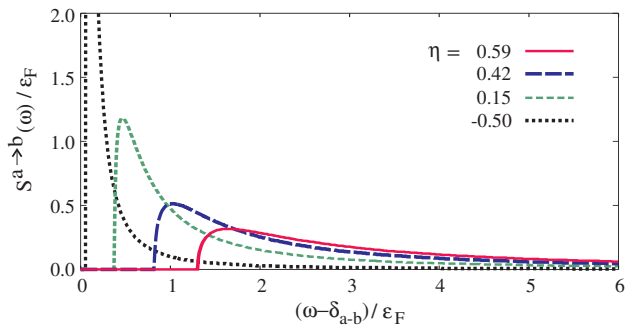}
   \caption{\label{fig:onech}
   Line-shapes for Type B transitions in
   the 2-level model. Type A transitions
   have zero strength in the 2-level model.
}
\end{figure}


The line shape of the response function $S^{i \rightarrow
j}(\omega)$ gives information about the many-body quantum state
which depend on the interatomic interaction. In current experiments,
the system is initially prepared by equally populating the two
lowest hyperfine states. However, subsequent variation of the
magnetic field, induces virtual scattering processes of particles
between all \emph{interacting} atomic states. Most of the available
experimental data, can be successfully interpreted in the framework
of the 2-level
model~\cite{legett,comte,rink,
ref:2level}, which only includes interactions between the lowest
hyperfine levels. Spin noise spectroscopy can access physics beyond
the 2-level model, and distinguish between
Fermi-Bose~\cite{holland,eddy} and multi-level Hamiltonian
models~\cite{ref:3level,bruun}.

A gas of $^{40}$K atoms ($I=4$) is an especially promising system
for studying models of the effective two-body interatomic
interaction~\cite{ref:3level}.
For $^{40}$K the subspace of interacting atomic states is very
restrictive, because the hyperfine coupling constant is negative so
that the lowest energy state is $m_F = - \frac{9}{2}$.
Experimentally, $^{40}$K atoms are trapped in the two lowest
hyperfine states, $|\frac{9}{2},-\frac{9}{2}\rangle$ and
$|\frac{9}{2},-\frac{7}{2}\rangle$. In the s-wave limit, the open
channel can couple to only \emph{one} closed-channel state
$|\frac{7}{2},-\frac{7}{2}\rangle$, and the interacting part of the
Hamiltonian reduces to a 3-level system. Figure~\ref{fig:k40_vert}
shows the schematics of the $^{40}$K structure, and the 3-level
Feshbach resonance model~\cite{ref:3level}. We denote the atomic
states $|\frac{9}{2},-\frac{9}{2}\rangle$,
$|\frac{9}{2},-\frac{7}{2}\rangle$, and
$|\frac{7}{2},-\frac{7}{2}\rangle$, as 1, 2 and 18, respectively. In
Fig,~\ref{fig:k40_vert} we also plot the expectation values of the
spin-projection operators for transitions in the 3-level model.


\begin{figure}[b!]
   \includegraphics[width=0.95\columnwidth]{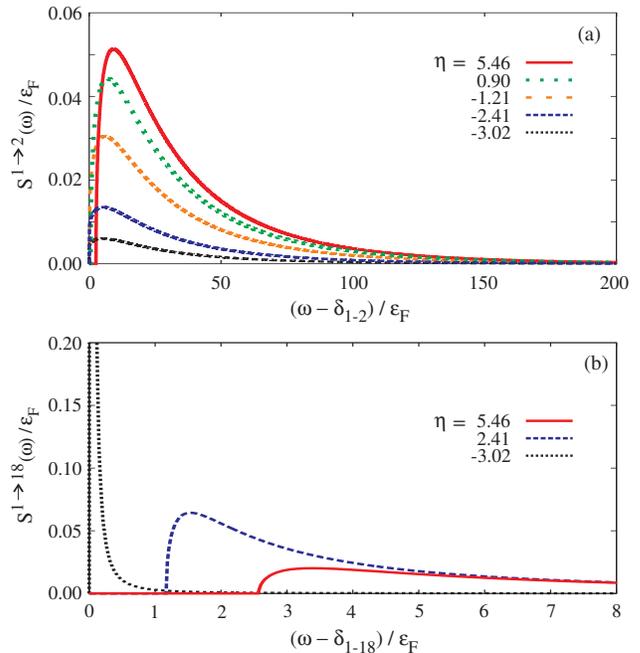}
   \caption{\label{fig:S3a}
   Line-shapes in the 3-level model
   for $^{40}$K. $S^{1\rightarrow 2}$ and $S^{1\rightarrow 18}$ are
   examples of Type~A and Type~B transitions, respectively.
   In the non-interacting limit, the lines of Type~B become
   $\delta$-functions while the lines of Type~A disappear. }
\end{figure}


\begin{figure}[b!]
   \includegraphics[width=0.9\columnwidth]{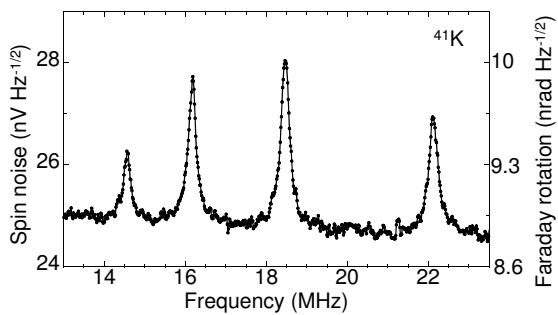}
   \caption{\label{fig:data}
   Spectra of spin noise, measured via Faraday rotation,
   in a $^{41}K$ vapor at 184~$^\circ$C (density N = 7.3 $\times$~10$^{13}$~cm$^{-3}$).
   Peaks correspond to $\Delta m_F=\pm 1$ transitions.
   The probe laser (4~mW) is detuned 100~GHz from the D1 transition
   (4$S_{1/2}$ - 4$P_{1/2}$; ~770~nm), and $B_{\perp}$ = 25.6 G. The vapor is contained
   in a 10~mm long cell, and the laser beam diameter (to 1/e peak intensity) is 65 $\mu$m.
   The noise data are shown in units of voltage ($\mathrm{nV}\, \mathrm{Hz}^{-1/2}$),
   and Faraday rotation fluctuations ($\mathrm{nrad}\, \mathrm{Hz}^{-1/2}$). The white noise floor is
   primarily from photon shot noise and amplifier noise.
   The discrete peaks show noise from electron spin
   fluctuations and have integrated values of
   1.3, 2.3, 2.6, and 2.1~$\mu$rad. Similar
   signal magnitudes are expected in ultracold atom systems (see
   text).
}
\end{figure}


In the 2-level model, the noise spectrum consists of lines
corresponding to occupied-to-occupied and occupied-to-unoccupied
states transitions. We use the Hartree-Fock-Bogoliubov (HFB)
formalism to calculate the zero temperature spin-spin response
functions $S^{i \rightarrow j}(\omega)$.  We find that the spin-spin
response function corresponding to a transition between two occupied
states is
\begin{align}
   S^{1 \rightarrow 2}(\omega)
   = &
   \frac{1}{\rho_0}
   {N(k_E)} \bigl [ (1 - \rho_{k_E}) \rho_{k_E} - \kappa_{k_E}^2
            \bigr ]_{E=(\omega - \delta_{12})}
\label{s12}
   \>,
\end{align}
where $N(k)$ is the density of states, and $\rho_k = \langle
a^\dag_k a_k \rangle$ and $\kappa_k = \langle a_k a_k \rangle$ are
the normal and anomalous densities, respectively. In Eq.~\eqref{s12}
we denote by $\delta_{12}$ the energy separation between levels 1
and 2. In the case of a transition from an occupied ($a$) to an
unoccupied state ($b$), the corresponding spin-spin response
function becomes
\begin{align}
   S^{a \rightarrow b}(\omega)
   = &
   \frac{1}{\rho_0}
   \bigl [ {N(k_E)} \ \rho_{k_E}
   \bigr ]_{E_k + (\epsilon_k - \mu) = \omega - \delta_{ab}}
\label{s13}
   \>,
\end{align}
where $\mu$ is the chemical potential such that $N_1 = N_2 =
\rho_0/2$. Since, for any $k$ value, $\rho_k^2 + \kappa_k^2 =
\rho_k$, the strength of the $1 \rightarrow 2$
(occupied-to-occupied, or Type~A) transition vanishes in the 2-level
model. The 2-level model for the line shape of the
occupied-to-unoccupied (Type~B) transitions depends on the
dimensionless parameter $\eta = (k_F a)^{-1}$ (see
Fig.~\ref{fig:onech}). Here $k_F$ is the Fermi momentum, and $a$ is
the s-wave scattering length. In the normal phase ($a \rightarrow
-0$), the line shape of a Type B transition becomes a delta-function
located at the energy separation between levels 1 and 2. On the BCS
side, the quasi-particle dispersion exhibits a local minimum at a
finite momentum value, and the corresponding singularity in the
density of states~\cite{ref:2level}, is reflected by the
characteristic shape of the spin-spin correlation function. On the
BEC side of the crossover, the singularity in the density of states
is located at zero momentum, and the spin-spin correlation function
has a smooth shape. The line shape of a Type B transition in the
2-level model provides an unambiguous signature for the presence of
fermionic superfluidity in the system.


The 3-level model involves the lowest two hyperfine levels, 1 and 2,
plus the topmost hyperfine state denoted by 18. In this model, the
particle number constraint becomes $N_1 = N_2 + N_{18} = \rho_0/2$.
Figure~\ref{fig:S3a} shows the characteristic shape of the lines
predicted by the 3-level model, as a function of the dimensionless
parameter $\eta = (k_F a)^{-1}$. The top panel illustrates the line
shape of Type~A transitions, such as $1\rightarrow 2$. The line
shapes for Type~B transitions are shown in the bottom panel. The
contrast between the predictions of the 2-level and the 3-level
model are evident: In the 2-level model, the transitions between the
occupied levels, $1 \rightarrow 2$, have zero strength independent
of $\eta$, while in the 3-level model the strength of the $1
\rightarrow 2$ transitions is \emph{nonzero}. The transitions $17
\rightarrow 18$, and $3 \rightarrow 18$ are allowed in the 3-level
model, but not in the 2-level model.

Because of the sum rule~\eqref{sum_rule}, the expected signal
strength for spin noise measurements in ultracold atom gases can be
estimated from corresponding measurements in classical alkali gases.
The theoretical results of Eq.~\eqref{power} apply to both classical
and quantum gases. The parameters in Eq.~\eqref{power} are the same
for the various isotopes of a given alkali atom.
Figure~\ref{fig:data} shows measured spectra of spin noise in a
vapor of $^{41}K$ atoms at 184~$^\circ$C. The discrete peaks,
readily visible above the white noise floor, result from electron
spin fluctuations. The integrated spin noise of the four peaks are
1.3, 2.3, 2.6, and 2.1~$\mu$rad, respectively. The ratios of the
measured integrated spin noise powers compare very well with the
theoretical results from Eq. (1). The overall magnitude of the
detected spin noise is a factor 2.7 lower than theoretical
expectation, which may result from uncertainties in the laser beam
diameter. In an ultracold gas of potassium atoms with a density of
10$^{13}$~cm$^{-3}$ and a trap length of 0.2~mm, we expect a similar
magnitude for the noise peaks with a 20~micron diameter laser
detuned 15~GHz from the optical resonance.  In ultracold fermionic
atom gases the frequency detuning can be significantly less than
15~GHz and thus much larger noise signals than shown in Fig. (4)
should be achievable.  We conclude that spin noise measurements in
ultracold gases of alkali atoms are feasible.


In summary, we have shown that electron spin noise spectroscopy can
be used to probe the quantum states of ultracold fermionic atomic
gases. The measurements proposed here have the unique ability to
unambiguously reveal the nature of the effective interactions
present in these systems, and to distinguish between various
many-body models for the quantum states of these systems. As a
consequence, the 2-level model predicts that transitions between
occupied atomic levels will have zero
strength~\cite{ref:instantaneous}. In contrast multi-level
Hamiltonians allow for such transitions. In particular, ultracold
$^{40}$K atom gases represent an ideal test case, because the
mechanism of the s-wave Feshbach resonance in this system consists
of only three interacting
atomic levels.\\


\textbf{Acknowledgments} This work was supported in part by the LDRD
program at Los Alamos National Laboratory. B.M. acknowledges
financial support in covering part of his travel expenses to
Cambridge through the ICAM fellowship program. We thank M.M.~Parish
and A.V.~Balatsky for valuable discussions.

Correspondence and requests for materials should be addressed to
B.M. (bmihaila@lanl.gov).



\begin{thebibliography}{99}

   \bibitem{anderson}
      Anderson,~M.H. Ensher,~J.R., Matthews,~M.R., Wieman,~C.E., \& Cornell,~E.A.,
      Observation of Bose-Einstein condensation in a dilute atomic vapor,
      \emph{Science} \textbf{269}, 198 (1995).

   \bibitem{davis}
      Davis,~K.B. \emph{et al.},
      Bose-Einstein condensation in a gas of sodium atoms,
      \emph{Phys. Rev. Lett.} \textbf{75}, 3969 (1995).

   \bibitem{bradley}
      Bradley,~C.C., Sackett,~C.A., Tollet,~J.J., \& Hulet,~R.G.,
      Evidence of Bose-Einstein condensation in an atomic gas with
      attractive interactions,
      \emph{Phys. Rev. Lett.} \textbf{75}, 1687 (1995).

   \bibitem{demarco}
      DeMarco,~B. \& Jin,~D.S.,
      Onset of Fermi Degeneracy in a Trapped Atomic Gas,
      \emph{Science} \textbf{285}, 1703 (1999).

   \bibitem{regal}
      Regal,~C.A., Greiner,~M. \& D.S.~Jin,
      Observation of resonance condensation of fermionic atom pairs,
      \emph{Phys. Rev. Lett.} \textbf{92}, 040403 (2004).

   \bibitem{bartenstein}
      Bartenstein,~M. \emph{et al.},
      Collective excitations of a degenerate gas at the BEC-BCS
      crossover,
      \emph{Phys. Rev. Lett.} \textbf{92}, 203201 (2004).

   \bibitem{zwierlein}
      Zwierlein,~M.W. \emph{et al.},
      Condensation of pairs of fermionic atoms near a Feshbach
      resonance,
      \emph{Phys. Rev. Lett.} \textbf{92}, 120403 (2004).

   \bibitem{pethick}
      Yakovlev,~D.G. and Pethick,~C.J.,
      Neutron star cooling,
      \emph{Annu. Rev. Astrophys.} {\bf 42}, 169 (2004).

   \bibitem{qgp}
      Meyer-Ortmanns,~H.,
      Phase transitions in quantum chromodynamics,
      \emph{Rev. Mod. Phys.} {\bf 68}, 473 (1996).

   \bibitem{holland}
      Holland,~M., Kokkelmans,~S.J.J.M.F., Chiofalo,~M.L., \&
      Walser,~R.,
      \emph{Phys. Rev. Lett.} {\bf 87}, 120406 (2001).

   \bibitem{eddy}
      Timmermans,~E., Furuya,~K., Milonni,~P.W., \& Kerman,~A.K.,
      Prospect of creating a composite Fermi-Bose superfluid,
      \emph{Phys. Lett. A} {\bf 285}, 228 (2001).

   \bibitem{ref:3level}
      Parish,~M.M., Mihaila,~B., Simons,~B.D., \& Littlewood,~P.B.,
      Fermion-mediated BCS-BEC crossover in ultracold K-40 gases,
      \emph{Phys. Rev. Lett.} \textbf{94}, 240402 (2005).

   \bibitem{bruun}
      Bruun,~G.M., Jackson,~A.D. \& Kolomeitsev,~E.E.,
      Multi-channel scattering and Feshbach resonances: Effective
      theory, phenomenology, and  many-body effects,
      \emph{Phys. Rev. A} \textbf{71}, 052713 (2005).

   \bibitem{ref:spin}
      Crooker,~S.A., Rickel,~D.G., Balatsky,~A.V., \& Smith,~D.L.,
      Spectroscopy of spontaneous spin noise as a probe of spin
      dynamics and magnetic resonance,
      \emph{Nature} \textbf{431}, 49 (2004).

   \bibitem{noise}
      Oesterich,~M., R\"omer,~M., Haug,~R.J., \& H\"agele,~D.,
      Spin noise spectroscopy in GaAs,
      \emph{Phys. Rev. Lett.} \textbf{95}, 216603 (2005).

   \bibitem{unitarity}
   Heiselberg,~H.,
   Fermi systems with long scattering lengths,
   \emph{Phys. Rev. A} \textbf{63}, 043606 (2001).



   \bibitem{legett}
      Leggett,~A.J.,
      Diatomic molecules and Cooper pairs,
      in \emph{Modern Trends in the Theory of Condensed Matter},
      edited by A. Pekalski \& R. Przystawa
      (Springer-Verlag, Berlin, 1980).

   \bibitem{comte}
      Comte, C. \& Nozi\`eres, P.,
      Exciton Bose condensation: the ground state of an
      electron-hole gas I.~Mean field description of a simplified
      model,
      \emph{J. Phys.} 
      \textbf{43}, 1069 (1982)

   \bibitem{rink}
      Nozi\`eres,~P. \& Schmitt-Rink,~S.,
      Bose condensation in an attractive fermion gas: from weak to
      strong coupling superconductivity,
      \emph{J. Low Temp. Phys.} {\bf 59}, 195 (1985).



   \bibitem{ref:2level}
      Parish,~M.M., Mihaila,~B., Timmermans,~E.M., Blagoev,~K.B. \&
      Littlewood,~P.B.,
      BCS-BEC crossover with a finite-range interaction,
      \emph{Phys. Rev. B} \textbf{71}, 064513 (2005).

   \bibitem{ref:instantaneous}
      B.~Mihaila \emph{et al.},
      Density and spin response functions in ultracold fermionic atom
      gases,
      \emph{Phys. Rev. Lett.} \textbf{95}, 090402 (2005).




\end{thebibliography}
\end{document}